\documentstyle[epsfig]{article} 

\begin{document}
  \title{ Quantum theory without Hilbert spaces}
 \author{C. Anastopoulos  \\ 
Department of Physics, University of Maryland, \\ College Park, MD20742, USA \\
email: charis@physics.umd.edu} 
\maketitle
 
\begin{abstract}
 Quantum theory does not only predict probabilities, but also relative phases for any  
experiment, that involves measurements  of an ensemble of  systems at different  moments of 
time. We argue, that any operational  formulation of quantum theory needs  an algebra of 
observables and  an object that incorporates the information about relative phases and 
probabilities. The latter is the (de)coherence functional, introduced by the consistent
 histories approach to quantum theory.  The acceptance of  relative phases as a primitive
 ingredient of any quantum theory,   liberates us from the need to use a Hilbert space and
 non-commutative observables.  It is shown, that quantum phenomena are adequately described 
 by a theory of relative phases and non-additive  probabilities on the classical phase space.
 The only difference lies  on the  type of observables that correspond to sharp measurements.
  This class of theories does not suffer  from   the consequences of Bell's theorem (it is not
 a  theory of Kolmogorov probabilities) and  Kochen- Specker's theorem (it has distributive 
"logic"). We discuss its  predictability properties, the meaning of the classical limit and
 attempt to see if it can be experimentally  distinguished from standard quantum theory. Our 
construction is  operational and  statistical, in the spirit of Kopenhagen, but makes
 plausible the existence of  a realist, geometric theory  for individual quantum systems.
 \end{abstract}

\renewcommand {\thesection}{\Roman{section}}   
\renewcommand {\theequation}{\thesection. \arabic{equation}} 
\let \ssection = \section \renewcommand{\section}{\setcounter{equation}{0} \ssection} 
 
 \section{Introduction}
 The first completed and consistent formulation of quantum theory was
 the Kopenhagen  interpretation. Its attitude was primarily operational: it considered quantum
 mechanics as a theory that  provides the  probabilities for measurement outcomes. Measurements are 
thought as the point of encounter between  the  classical world of apparatuses  and observers 
and  the  microscopic world of atoms. All predictions were phrased in terms of  ensembles and
 ensemble averages. Kopenhagen would not deal with individual systems.  

Such a description, however successful, did not answer the important question: how is an 
 individual  system described and what are its properties? This was Einstein's  demand 
for "elements of reality" and Bohm's later emphasis on the need for ``ontology in quantum 
theory".  This question is of utmost  importance  in physics:  in spite  of the remarkable
 success of quantum mechanics,  it was thought imperative, that new frameworks  have to
  be devised in order to  address it.

  There were  two main directions: either supplement quantum theory with extra variables that
 would  describe individual systems ("hidden variable theories"), or keep the existing 
formalism and try to  interpret it in a {\it realist} sense, as though it refers to properties 
of  individual systems (state vector reduction \cite{vNeu}, 
relative state formulation \cite{Eve, Gra73},  consistent histories 
\cite{Gri84,Omn8894, GeHa9093,Har93a}...).  Both approaches come  to insurmountable obstacles
 due to two theorems: Bell's \cite{Bell64} and Kochen and Specker's \cite{KoSp67}.  

 Bell's theorem and the subsequent experiments \cite{Asp82} prevent hidden variable theories 
from being local, while the Kochen-Specker's theorem forbids realist theories  from asserting 
the existence of  definite (i.e. non-contextual) properties for  individual  physical  systems.
 Hence from the first category only non-local theories, such as the precious but inelegant  Bohmian mechanics 
\cite{Bohm52, BoHi} have survived. From the second one, all approaches  have to accept that  
properties of quantum  theory are contextual - something very disturbing for any theory that
 pertains to "objectively"  describe physical phenomena. It would, then, be fair to say, that 
seventy two years later,   the Kopenhagen interpretation has survived  all assaults. This is 
due to  its balance and moderation: it claims little and does not refrain from admitting 
ignorance (but will  not  admit   incompleteness, either).  

  Kopenhagen quantum theory is essentially a model for describing experiments and is based on 
the  notion of probability. We count the number of events in an ensemble of physical systems 
and define probabilities for  these events from   their relative frequency. But counting  
occurences  of events does not exhaust the physical or  observable content of quantum theory. 
 Relative phases are observable, as was first shown by  Bohm and Aharonov \cite{BoAh59} and 
these phases cannot be  solely described by probabilistic concepts.  What is more, the 
consistent history approach has emphasised, that  when we examine properties of systems at 
more than one moments of time, we cannot use  standard probability theory. The physical 
probabilities correspond to a  non-additive measure. This  non-additivity comes  from the
 presence of interference phases. In fact, these are of the same origin as the Bohm-Aharonov 
 phases; they are all generated by the Berry connection \cite{AnSa00}.   

 The relative phases are important and ever present; they are also measurable. This is what  we show in  detail in 
section 2. All possible measurement outcomes, whether of phases or of probabilities  can be 
encoded in an object, that was first  introduced in the consistent histories approach: 
 the decoherence functional.   We further argue that one should {\it not}  think of quantum 
theory  as a probability theory. Probability theory, its axioms  and its concepts are 
mathematical constructs;  there is  no {\it a priori} reasons for them to describe physical
 reality. One has to provide {\it physical}  arguments, before any mathematical model is
 chosen to be applied in a physical situation. In particular, an operationalist attitude to 
 probability has to admit that certain mathematical axioms (additivity of probabilities) are
 not  warranted by any operational procedure.  

Quantum theory  is not adequately described {\it solely}  in terms of
 probabilistic concepts. A theory for the quantum phenomena ought to provide values  
for both   "probabilities" and relative  phases.  Once we admit this, there is no reason to
 be constrained by the Hilbert space formalism. All  that is needed for an operational 
formulation of a quantum theory is   an algebra of observables and the decoherence functional, 
which contains the  information about all possible measurements. (We think, that a more 
appropriate name for the functionality  of the latter, would be the  coherence functional.)  
 The algebra of observables does not have to be  non-commutative. In fact, as  we show in 
section 3, one can consider that our observables are functions  on the classical phase  space 
and use the Wigner transform to  construct a theory that fully reproduces the  predictions of 
quantum theory. Its only  disagreement is in what each theory considers as sharp measurements:
 in our  theory sharp measurements  correspond to subsets of a space $\Omega$ (the phase space),
 while in standard  quantum theory  they are projection operators.
 The quantum behaviour is all  contained in the relative phases of the coherence functional.  

We are careful to define our theory as an operational one, in the spirit of Kopenhagen. 
But it  easy to see that any theory starting therefrom would be able to sidestep both Bell's 
and Kochen-Specker's theorems. We avoid the former, because we do not have a probability 
theory - our primitive  concepts are intensities and relative phases. And we avoid the latter,
 because the "logic" of the theory is distributive, as a corollary of the commutativity of the
 algebra of observables.  Classical probabilities and, as a limiting case, determinism,  are 
obtained as approximate theories. In some regime,  we can describe our experiments using 
classical probability theory (if we coarse-grain enough to suppress  the relative phases)
 and if one wishes one might use rules and interpretation of classical probability, in order
 to   to make predictions. These are not unambiguous, but at least clearer than the  quantum 
ones. 

 This theory is operationally equivalent to quantum theory. But it may lead to more than that.
 Having  abandoned the Hilbert space one has much more freedom in trying to construct a
 quantum theory  for individual systems. In fact, one has the freedom to look for a geometric
 origin of quantum phenomena, and exploit all the technical machinery of the 
{\it quantisation} approaches in doing so. This is something  suggested by some recent results
 in the consistent  histories programme \cite{AnSa00,Sav99a} . The ideal result  of such an 
attempt  would be to understand the physical reason, why  relative phases appear and how  they
 are related to statistics.  

\section{ Filter measurements}

 Our first aim is to recover the Kopenhagen quantum theory. We, therefore,  take a minimalist
 stance towards what a physical theory should do. We just  demand that it can provide an 
adequate model for describing our possible experiments.   For this reason we find convenient
 to idealise experiments in the fashion of von Neumann  and Jauch 
\cite{Jau}
or the operational approach to 
quantum theory \cite{Dav, BGL}. We  start by stating our
 basic operational concepts. 

   First, we have   the {\it sources} $S$: they  prepare an
 ensemble  of physical systems (which we will call {\it particles}). This ensemble we  will call  a
  {\it beam}. By its {\it intensity} we  mean the number of particles it contains, which is 
assumed to be as large as we want.  An experiment that can be modeled corresponds  to a beam 
 passing  through a  sequence of {\it filters}: these are experimental  setups,   that allow a
 particle  to pass only if it is found to satisfy a certain property.  The filter then 
typically reduces the intensity of a beam.  We choose to idealise measurements by filters
 rather than by pointer devices, because the  latter are more difficult to visualise when we
 have successive measurements. A pointer  device can be viewed as a collection of mutually 
incompatible filters, placed at (roughly)  the same point in the path of the beam.  The result 
of a beam passing through a series of  filter we shall call an experimental history or short
 a {\it history}. 

 We will also assume that we have some way of determining the intensity of each beam, by 
 measuring the number of particles that were incident on  a {\it detector}. The relative  
intensity of the beam  after passing through a number of filters $C, D, \ldots, E$, we will 
call the  {\it probability} determined through this experiment. No mathematical meaning is
 to be  given to this word: in particular it does not refer to Kolmogorov probability. It 
 denotes the  ratio of intensities, and the choice for this word  implies the  frequency
 interpretation of probabilities, which is the natural in an operational setting. 

 We will
 also assume that we have devices, such as beam splitters and beam recombinators. 
We also  have screens, upon which we can see  interference patterns. 

 The experiments idealised here refer to ensembles rather than individual particles.  A 
non-deterministic theory can only model ensembles, at least when no further information  about
 the physical content of the system is assumed. One model might be able to make  predictions
 about the individual system, if we decide to push the frequency interpretation  of probability
 to its limits. Nonetheless, this is not necessary and we should attempt  this only if we have
 a good grasp of the meaning of the laws that govern the measurement  of beams.   

 \subsection{Classical beams}  
Let us consider  experiments, where the beam is assumed to satisfy classical probability
  theory. The physical system is characterised by  a number of parameters that determine 
 points on a space $\Omega$.  Then the beam will be  described by a normalised, positive function on
 $\Omega$,  $\rho(x)$, $x \in \Omega$. Each filter will be described by a  function 
$\chi(x)$, that  truncates all values of $x$ outside the range that characterises the filter.
 A perfect  filter has for $\chi$ the characteristic function of a subset of $\Omega$. An
 imperfect  filter does not perform according to its specification all the time, it should
 therefore  be described by a smeared characteristic function. We shall write as $C(x)$ the
 characteristic  function of a subset $C$ of $\Omega$.   

Consider a particular two-filter experiment as in figure 1.  A beam $\rho$ leaves the source 
and passes through two  filters $C$ and $D$  at times $t_1$ and $t_2$ respectively. After 
passing $C$ the beam has become $ C \rho$  and after $D$ it is $D C \rho$. The detector $I$
 then measures the intensity  of the beam which is equal to $\sum_x (DC \rho)(x)$. This way 
we measure the  probability for the history ``$C$ and then $D$". This we shall denote as 
$p(C,t_1;D,t_2)$.

\begin{figure}
\centerline{ \psfig{file=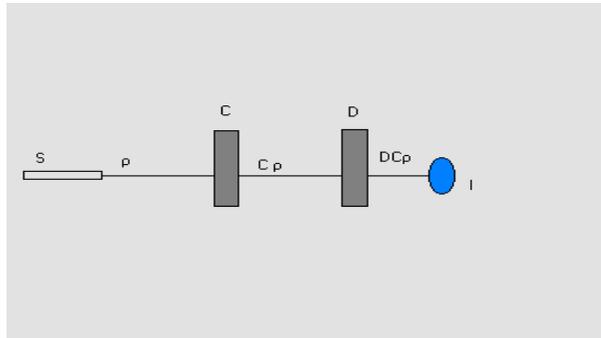,width=8cm, height=4.5cm, angle=0} }
\caption{ An experiment with classical beams. The beam described by $\rho$,
leaves the source $S$, passes through the filters $C$ and $D$ and its intensity is measured 
by the detector $I$.} 
\end{figure}

 Note, that we have assumed that the probabilities for the beam have no  self-dynamics.
  
 Suppose we carry these experiments with many different filters.
 In particular,  we can consider
 a class of filters $C_i$, each of them corresponding  to a value $\lambda_i$ of an
 observable $A$  \footnote{The determination  of an  observable, that  corresponds to a given 
set of filters  is an important fact that a theory describing the physical system has to 
provide.  In practice, it is determined by reference to other physical systems and can be
 argued to be  eventually tied to measurement of time or space, or to the counting of 
numbers.}.    Then, to this observable we can assign a function 
 \begin{equation} 
A(x) = \sum_i \lambda_i C_i(x) 
\end{equation}

 Suppose we perform a number of experiments for all possible filter  configurations 
``$C_i$ and then $C_j$'' and that we measure the resulting quantities 
$p_{ij}= p(C_i,t_1;C_2, t_2)$. Then the correlation function for the observable $A$ can be
 reconstructed as  
 \begin{equation}
 <A_{t_1} A_{t_2} > = \sum_{ij} \lambda_i \lambda_j p_{ij} 
\end{equation} 
 In general, the information for all possible measurements at all possible times is encoded 
 in the stochastic probability measure $d \mu(x(\cdot))$, on the space of all paths  
$x(\cdot)$ from ${\bf R}$ to $\Omega$. From this, we can predict or reconstruct  any 
probability or correlation function of the theory. Conversely, a sufficiently large  number 
of beam experiments suffices to reconstruct (with any desired accuracy) the  stochastic
 probability measure.  

\subsection{Quantum beams: intensities} 
 Let us now try to repeat the same type of experiments in the quantum case. To a beam
 we  associate a vector $|\psi \rangle$ in a Hilbert space $H$. According to the rules of
 quantum  theory a filter will correspond to a projection operator $P$.   Let us then put
 two filters $P$ and $Q$ from which the beam will pass, at time $t_1$ from  $P$ and at 
time $t_2$ from $Q$. Then after passing from $Q$ the beam will be described by  the vector
 $| f \rangle = QP | \psi \rangle $.

\begin{figure}
\centerline{ \psfig{file=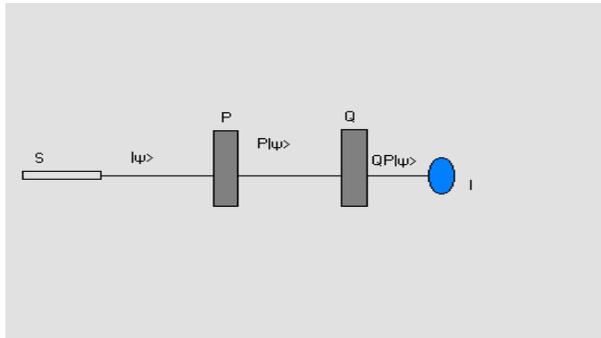, width=8cm, height=4.5cm, angle =0}}
\caption{ Measuring intensity of quantum beams. The beam $|\psi \rangle$ leaves the source $S$,
passes through filters $P$ and $Q$ and its intensity is measured at $I$. The setup is identical to the one for 
the classical case.}
\end{figure}

     From the rules of quantum theory we know that the relative intensity of this beam  will 
be 
 \begin{equation} 
\frac{\langle f| f \rangle}{\langle \psi | \psi \rangle } =  
\frac{\langle \psi | PQ QP| \psi \rangle}{\langle \psi | \psi \rangle }
 \end{equation} 
 This gives the probability $p(P,t_1;Q,t_2)$ that the history  "first $P$ and then $Q$" will 
be realised.   Now consider the case of  three distinct experiments. In the first we take 
the  first filter to be $P_1$ and the second $Q$. In the second, we replace $P_1$ with  $P_2$,
 where $P_2$ is another filter such that $P_1 P_2 = 0$ (no beam can pass those  two filters in 
succession) and $P_1 + P_2 = 1$ (there does not exist any other filter  $R$ such that both 
$P_1 R = P_2 R = 0$). In the third experiment  we put no filter before  $Q$. We can then 
measure the beam intensities in three cases and when we compare them  we find that in
 general  
\begin{equation}
 p(P_1,t_1; Q,t_2) + p(P_2,t_2;Q,t_2) \neq p(Q,t_2)
 \end{equation} 
  This means that we cannot use standard probability theory, in order to construct  a model
  that predicts  the beam intensities. The Kolmogorov additivity condition that a probability
  theory has to satisfy fails.   

In general, the correct formula for the intensities would have to take into account the  self-dynamics 
for the system that constitutes the beam. This 
would be generated by a  Hamiltonian operator $H$ and would give for the vector $| f \rangle$,
 that describes the  beam incident on the detector at time $t_f$
 \begin{equation} 
| f \rangle = e^{-i H(t_f - t_2)} Q e^{-iH(t_2 - t_1)} P e^{-iHt_1} | \psi \rangle
 \end{equation}

 This gives the following expressions for the intensity of the beam after passing through  
two filters 
 \begin{equation} 
p(P,t_1;Q,t_2) = \langle \psi | e^{iHt_1} P e^{iH(t_2-t_1)} Qe^{-iH(t_2 - t_1)} P e^{-iHt_1}| 
\psi \rangle 
\end{equation} 
If  we restrict to filters $P_i$, each corresponding to a value $\lambda_i$ of an  observable
 $A$, then we can write the statistical correlation function for $A$ 
\begin{equation} 
<A_{t_1} A_{t_2}> = \sum_{ij} \lambda_i \lambda_j p(P_i,t_1;P_j,t_2)
 \end{equation} 
This is a real number: it is {\it different} from the quantum mechanical correlation  
functions, which are complex valued.  
\begin{equation}
 G_A(t_1,t_2) = \langle \psi|e^{iHt_1} A e^{-iHt_1} e^{iH t_2 } A e^{-iH t_2} 
| \psi \rangle 
\end{equation}
 Clearly this correlation function, cannot be determined  by measurement of intensities as 
described so far.  

 \subsection{Quantum beams: relative phases} 
Quantum beams contain more information than their intensities. Their relative phase is also 
 of physical interest.   In other words, when a beam passes through a succession of filters, 
it undergoes more changes,  than the ones encoded in its intensity. And these changes are
 measurable only through  procedures of combination of beams and comparison of beams. 
In effect if the beam passes  through the filters $P$ and $Q$ the (complex) quantity  
$ \langle \psi|QP| \psi \rangle $ can be experimentally determined. 

 Let us give a hypothetical example of how to measure this phase structure. First, let us
 assume  that each source $S$ that prepares a beam $| \psi \rangle$ comes from its
 manufacturer  together with a set of fine filters labeled $| \psi \rangle \langle \psi |$.  
When the beam exits the second filter $Q$ it enters this fine filter becoming 
 \begin{equation}
 \langle \psi | Q P | \psi \rangle | \psi \rangle 
\end{equation} 

The measurement of the relative intensity of this beam gives as an outcome a  positive number 
 less than one 
 \begin{equation}
 r(P,t_1;Q,t_2) = |\langle \psi | Q P | \psi \rangle| 
\end{equation}

\begin{figure}
\centerline{\psfig{file=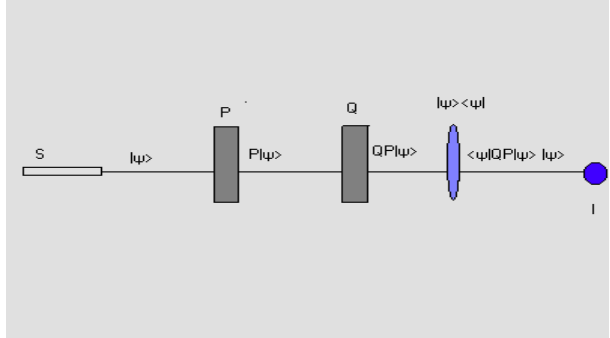, width=8cm, height=4.5cm, angle=0}}
\caption{ Determining $|\langle \psi|QP| \psi \rangle|$ through measurement of 
intensities. A beam $|\psi \rangle$ leaves the source $S$, passes through filters $P$ and $Q$ and 
then through $| \psi \rangle \langle \psi |$, before it is measured at $I$.}
\end{figure}

 But we can also perform another measurement. Let us consider a source emitting a beam 
 $|\psi \rangle + | \phi \rangle$. And let us assume we can monitor its wave pattern at  a
 screen SC and store it in memory.   Then we carry another experiment where
 $|\psi \rangle + | \phi \rangle$ passes through  a beam splitter and splits into its 
$| \psi \rangle $ and $| \phi \rangle$ component.  The component $ | \phi \rangle$ is kept as a 
reference beam, but the $|\psi \rangle $  component  has to pass through filters $P$, $Q$ and 
$| \psi \rangle \langle \psi |$ as before. The resulting beam 
 $ \langle \psi | Q P | \psi \rangle | \psi \rangle $ is recombined with $| \phi \rangle$ and
 we can see  its interference pattern on the screen. When we compare this interference
 pattern with the previous one,  we notice a phase shift equal to the  phase of 
 $\langle \psi | Q P | \psi \rangle $. This we will call
 $e^{i \theta(P,T_1;Q,t_2)}$  \footnote{The comparison of interference patterns is, for
 instance, the way the Bohm-Aharonov phase was originally measured \cite{Cha60}. It should be
 remembered, that the measurement  of phases is not something that is adequately described by
 the practical rule that "measured  quantities correspond to self-adjoint operators". This
 point was, in fact, the original motivation  for Bohm and Aharonov's work. The relative phases
 are obtained by comparing patterns, rather than  reading pointers in devices. In a sense, a
 phase measurement  is  a non-local way  of extracting information from a quantum system.}. 
It is, in fact, a generalisation of Berry's phase \cite{SaBh88,AnSa00}.

\begin{figure}
\centerline{\psfig{file=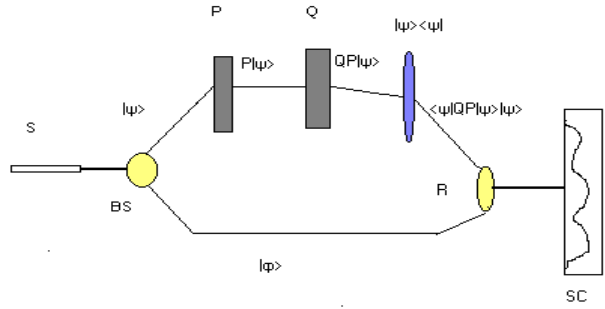, width=8cm, height=5cm, angle=0}}
\caption{ Measuring the relative phase of the beam. A source $S$ emits a beam 
$| \psi \rangle + | \phi \rangle $, which a beam splitter splits into $\psi \rangle$ and $| \phi \rangle$.
The component $| \psi \rangle $ passes through filters $P$ and $Q$ and then through $ | \psi \rangle 
\langle \psi | $, before it is recombined with $ | \phi \rangle $ and their interference pattern measured 
at the screen $SC$. Comparing this pattern, with the pattern of $| \psi \rangle + | \phi \rangle $ 
(which we have from a previous experiment) enables the determination of the relative phase between 
the history with filters $P$, $Q$ and the history with no filter, for the beam $ | \psi \rangle$.  }
\end{figure}

  There are  more efficient ways to measure the complex number  
$ \langle \psi |QP| \psi \rangle$, without needing the fine filter  
$|\psi \rangle \langle \psi |$. We can for instance consider it as correlated with another 
system, which is suffiently controlled,  and find the relative phase by the change in  certain 
transition rates. This is how the Berry phase was measured in  NMR
 interferometry \cite{SMP87}. Or perhaps the phase could be obtained  by studying interference 
of the beam $QP | \psi \rangle$ with a large number of reference beams  $| \phi_i \rangle$. 
  The exact procedure of measurement  makes little difference in the conceptual description,  
though. The important statement is that, {\it any experiment that measures phases must  
necessarily  compare or interfere two beams}.

 This means that the phase is not absolute; 
 therefore, it  cannot be read from intensity measurements of a single beam.  In fact we can
 find the relative phase between two beams that have passed through different  filters: 
it can be read from the comparison of their interference patterns with a reference beam  
$| \phi \rangle$.   This phase is equal to the one we would measure for a single beam passing 
 successively through $P_1,Q_1,Q_2,P_2$, i.e. adding the filters of the second beam in 
 reverse order. (If $H \neq 0$, the placement of the filters is also important).  
 As far as the correlation functions are concerned, we can easily see that the quantum  
mechanical correlation functions \footnote{ To the best of my knowledge, the  only 
experimental determination of correlation functions in quantum theory is in the case of the
 electromagnetic field.  There, correlations of photon number  can be  measured. But, since for
 the EM field, the photon number commutes with the Hamiltonian, the quantum  mechanical and 
the statistical correlation functions coincide and are real-valued.}  can be measured as 
 \begin{equation}
 \langle A(t_1)A(t_2) \rangle = \sum_{ij} \lambda_i \lambda_j [re^{i \theta}](P_i,t_1;P_j,t_2) 
 \end{equation} 

 \section{Models for quantum theory} 
How do we encode the results of filter measurements? Let us denote by $\alpha$ the  series of 
filters $P_{t_1}, \ldots, P_{t_n}$. We can  define the operator 
\begin{equation} 
C_{\alpha} = e^{iH t_1}P_{t_1}e^{-iHt_1} \ldots e^{iHt_n}P_{t_n} e^{-iHt_n}
 \end{equation}
  For a pair of histories $\alpha$ and $\beta$, we can write the object
  \begin{equation} 
d(\alpha, \beta) = \langle \psi|C^{\dagger}_{\alpha} C_{\beta} | \psi \rangle = 
Tr (\rho C^{\dagger}_{\alpha} C_{\beta} )
 \end{equation} 
If $\alpha \neq \beta$ this gives the relative phase between the two  histories. If 
$\alpha = \beta$, this gives the relative intensity of the  final beam. 

 \subsection{The consistent histories approach} 
This object was introduced by Gell-Mann and Hartle in the context of the consistent histories 
 approach to quantum theory. They called it  the decoherence functional. Its function is to 
determine, when classical probability can be  used to describe a quantum system. 

 The
 consistent histories is a realist formulation of quantum theory for  individual systems. Its 
main tenet is that probabilities can be defined in a consistent set,  i.e. a set of exhaustive 
 and exclusive histories that satisfy  
\begin{equation}  
d(\alpha, \beta) = 0,
  \end{equation}
 for $\alpha \neq \beta$.  Propositions in this set can then be described by classical logic 
 and one can make predictions and retrodictions within this set. But all such predictions  
are contextual: they have to make reference to a given consistent set. Otherwise, one  might 
obtain contradictory inferences \cite{DoKe96,Kent97}. This pathology is typical in all 
realist  interpretational schemes for quantum theory: it is of the same nature as the
 Kochen-Specker paradox \cite{Ish97} \footnote{  We should remark that this pathology does  {\em not} lie 
at the level of logical consistency of the theory  \cite{GriHa98, Gri99}, but at the epistemic level, since 
it forces   a redefinition of fundamental notions ({\em truth} or {\em property of a system}), 
which is much weaker than what is employed in scientific practice.}. 

All treatments of consistent histories in the literature focus on its status as a realist interpretation.
However, the formalism makes sense even if it is interpreted solely as an operationalist one, i.e.
a generalisation of the Kopenhagen interpretation dealing with time-ordered series of measurements.
For this reason we  make a distinction 
 in this paper, between the histories formalism and the consistent histories approach, which is 
the realist interpretation of the formalism in terms of consistent sets.

The consistent histories interpretation ignores the
 values of the off-diagonal  elements of the decoherence functional. It focuses on the 
probability aspect of quantum theory  (only intensities) and ignores the phase aspect, even if 
there exist meaningful operational  schemes of measuring them. This we think is a severe 
omission: not because it affects the  logical consistency  of the theory.  The consistent
 histories  framework   claims that it describes adequately  all measurement situations. 
The point is, rather, that it contradicts the motivation of the scheme as  providing a  
 realist description of physical phenomena: The reasonable expectation one would have for  
a realist formulation of quantum theory {\it is that it should explain the predictions of  
ensemble measurements, in terms of properties of the individual systems.} 

The consistent histories  scheme, as any scheme that is based solely on probabilities, cannot 
do that for the relative phase measurements.    An  insistence that only probabilities are 
physically relevant, inevitably    gives physical meaning to only   a  part of the information
 contained in the decoherence functional \footnote{To be precise, if  we are to stay purely 
at the level of probabilities,  the  real part of the decoherence functional  is  sufficient 
and necessary, since it contains the information about the non-additivity  of the probability 
measure: $2 Re d(\alpha, \beta) = d(\alpha + \beta, \alpha + \beta)  - d(\alpha,\alpha) -
 d(\beta, \beta)$.   In this sense, probabilities are exactly half  the content of the 
decoherence functional. Nonetheless,  Gell-Mann  and Hartle have  identified the full
 decoherence functional as the  physically  relevant object, using arguments based on 
persistency of records.}.  

\subsection{The  filters' algebra} 
 Our stance, so far, is  operational: filter measurements allow us to determine  (in principle)
 all possible values of the decoherence functional . As such it is  the object that 
incorporates all information about measurements, as the stochastic  measure does in the case 
of classical beams.  One could still question, whether there is any more information we can 
get through a  particular implementation of our filter-measurement scheme. Is it possible that
 the  comparison of three histories  gives results that are not contained in the measurement 
 of intensities and comparison of two histories?   Sorkin has demonstrated that in standard 
quantum theory the answer is negative \cite{Sor94,Sor97}. All  information obtained from the 
comparison of three histories is entirely attributed to the  interference between all pairs of
 them. And this stems from the fact that quantum theory  is based on the complex 
numbers rather than any other algebraic field. 

  We are then interested in finding a
 mathematical model that describes this set of experiments.  What would be the possible 
ingredients?  First we need to identify the mathematical objects that correspond to filters 
$P$. The simpler condition is to consider that  the filters correspond to   idempotent
 members  of an associative algebra ($P^2 = P$).    We shall denote the algebra by ${\cal A}$ 
and by $I({\cal A})$ the set of its  idempotent elements. The algebra has to have a unity $1$,
 which corresponds to no filter   and a zero $0$, which is the completely opaque filter. 
There is then the following correspondence between mathematical objects and physical 
operations.  
\begin{list}{$\bullet$}
 \item   The idempotency condition $P^2 = P$ means  that the  (almost instantaneous)
 succession of two identical filters changes the beam in an identical fashion    as one single
 filter does. 
 
\item  The condition $P_1 P_2 = P_1 $ implies that the filter $P_1$ is finer (or more 
restrictive) than the filter $P_2$.  (Conversely $P_2$ is {\it coarser} than $P_1$). This is 
represented  as $P_1 \leq P_2$, and defines a partial ordering in the space of filters.  

\item   The condition $P_1P_2 = 0$ corresponds to two incompatible filters: if we put  them in
 succession they act like the opaque filter. 

  \item   For incompatible  filters $P_1 + P_2$ denotes a filter that is coarser than $P_1$ 
and  $P_2$ and we cannot construct any finer filter that has this property.    

\item   If the algebra is  not Abelian,   neither addition nor multiplication of generic
 filters  in the algebra have a natural interpretation in terms of simple operations. 

  \item    If among a class of filters $P_i$, each of them  corresponds to a value $\lambda_i$ 
of an observable $A$, this observable is represented as  an element $A = \sum_i \lambda_i P_i$ 
of the algebra ${\cal A}$.  
\end{list}

   An experiment consisting of a sequence of filters $P_1, \ldots P_n$ ,   placed at moments 
 $t_1, \ldots, t_n$, can be described by the element $P_1 \otimes \ldots \otimes P_n$ of the
 {\it tensor product } algebra $ \otimes_i {\cal A}_{t_i}$,  where ${\cal A}_{t_i}$ is a copy \
of the filter algebra labeled by time $t_i$.  Elements of this tensor product algebra can be 
viewed as corresponding to (possibly)  time-averaged observables for  this system. The algebra 
${\cal A}_h =\otimes_t {\cal A}_t$ includes all possible filter  sequences $\alpha$ we can
 construct for the system \footnote{ The tensor product over $t$ of the filter algebras can 
have many interpretations. We can consider it as a space containing all {\it finite} tensor
 products of algebras ${\cal A}$ \cite{I94},  or a genuine  tensor product over all values of
 $t \in {\bf R}$ (which makes the algebra too large), or a construction of  an object that 
resembles a tensor product over a continuous variable \cite{IL95,An00b}.  
For simplicity we shall consider that $t$ takes values in a finite set with a large  number of 
elements. As such there is no problem in the construction of this object. }.  

 We shall call 
any finite filter sequence a {\it history }:  it will be represented by small greek letters. 
As a mathematical object it is represented  by an element of ${\cal A}_h$ {\it that can be 
written as a tensor product  of idempotent  elements of ${\cal A}$}, i.e. $\alpha = P_{t_1} 
\otimes P_{t_2} \otimes \ldots \otimes  P_{t_n}$ \footnote{ The  use the tensor product was 
proposed by Isham \cite{I94} in the  context of a temporal quantum logic reformulation of 
the consistent histories approach. The constructions presented in sections 3.2 and 3.3  are  a
 rephrasing of this quantum logic scheme in an operational language.}.

 The moments of time
  $\{t_1, \ldots, t_n \}$ upon which a filter has been set defined  the {\it temporal support}
 of this  history.   Two histories are {\it incompatible}, if at some  time 
$t \in T_1 \cap T_2$ the  corresponding filters are incompatible.   Consider two histories
 $\alpha_1$ and $\alpha_2$ have temporal supports $T_1$ and $T_2$. We will, then, denote by 
$\tilde{\alpha}_1$ and $\tilde{\alpha}_2$ the same histories,  but viewed as having temporal 
support $T_1 \cup T_2$ (we just append the trivial filter at  their non-common points). 
A history $\alpha$ is {\it finer} than $\beta$ (denoted $\alpha \leq \beta$ ), if  at all
 $t \in T_1 \cup T_2$ the filters of $\tilde{\alpha}$ are finer than the filters  
of $\tilde{\beta}$. 

     Two incompatible  histories with the same temporal support can be added to give  an idempotent
 elements of ${\cal A}_h$, but it is not necessary that they form a history (i.e an idempotent
 element of the form $P_1 \otimes \ldots \otimes P_n$). In the consistent histories scheme, 
they are said to  correspond to {\it propositions about the possible outcomes of the
 experiment}. A case where  $\alpha + \beta $ corresponds to a history is the following: 
when $\alpha$ and $\beta$ are 
 constructed from   filters that are at all time-points, but  one, in which their filters 
are incompatible, e.g. if  $\alpha = P_1\otimes Q$, $\beta = P_2 \otimes Q$, with
 $P_1P_2 = 0 $, then $\alpha + \beta = (P_1 + P_2) \otimes Q$ corresponds to a history. 
This would not be true for $\alpha = P_1 \otimes Q$ and
 $\beta = P_2 \otimes Q'$, with $[Q,Q'] \neq 0$.
 Another case is when the two histories are disjoint because they have disjoint 
temporal supports.
This is,  for instance,  the case of the history $\alpha = P_{t_1}$ and the history $\beta = Q_{t_2} \otimes Q'_{t_3}$, with 
(say) $t_1 < t_2 < t_3$. Then we define the "addition" as $\alpha + \beta := P_{t_1} \otimes Q_{t_2} \otimes Q_{t_3}$ is another 
filter history on the Hilbert space $H_{t_1} \otimes H_{t_2} \otimes H_{t_3}$.
 We
 shall call two filter histories $\alpha$ and $\beta$ {\it operationally additive}, if $\alpha + \beta$ 
is a filter history.   

 If we want to consider experiments carried out simultaneously on two different physical
  systems, each characterised by a filter algebra ${\cal A}_1$ and ${\cal A}_2$, the filter 
 algebra for the total system is ${\cal A}_1 \otimes {\cal A}_2$. This tensor product  is
 distinct from the one used earlier to construct a space, where possible histories might  be
 embedded.   

 \subsection{The coherence functional}
  To each source we then assign an object $d$, which is a complex valued map of pairs of  
measurements. It incorporates all information about all possible  filter
 measurements for this beam.   Unlike the consistent histories scheme, we want to emphasise 
the overall importance of the  relative phases. We think it would be more precise in our 
scheme to call this object  the {\it coherence functional}. Its diagonal elements 
$d(\alpha, \alpha) $ measure the relative intensity of  the final beam, while its off-diagonal
 the relative phase between the beam having passed  through different series of filters. 
As such it can be extended to a functional over  ${\cal A}_h \times {\cal A}_h$. 

 Quantum theory suggests that this functional ought to satisfy the following properties
 \cite{IL94} 
\\ \\ 
1. {\it Positivity} : $d(\alpha, \alpha) \geq 0$; all beams have positive intensity. 
\\ \\ 
2. {\it Normalisation} : $d(1, 1) = 1$; in absence of filters the intensity does not change.
 \\ \\
 3. {\it Hermiticity} : $d(\alpha, \beta) = d^*(\beta, \alpha)$; the reverse order of 
comparing two histories, gives the opposite number for the relative phase. 
\\ \\ 
4. {\it Additivity} : $d(\alpha+ \beta, \gamma) = d(\alpha,\gamma) + d(\beta, \gamma)$, if 
$\alpha$ and $\beta$ are disjoint histories; this is suggested by the corresponding structure
 in quantum theory. 
 \\ \\
 5. {\it Triviality} : $d(0, \alpha) = 0$;  from an opaque filter no beam can pass. 
\\ \\
  These conditions on the coherence functional are natural consequences of the operations we
  aim to describe. Only number 4. is a mathematical assumption, that is not intuitively 
 evident, but clearly suggested as fundamental from quantum theory.  Note that the histories 
$\alpha$, $\beta, \alpha + \beta, \ldots$ in properties 1-5 refer to idempotent  elements of ${\cal A}_h$ of
 the form $P_1 \otimes\ldots \otimes P_n$, that correspond to  filter measurements. 
But they hold for general indempotent elements  of ${\cal A}_h$. There exist,  though, 
 properties of the  coherence functional in standard quantum theory, that cannot be extended
 this way \cite{IL94}.  The most important are 
\\ \\
 6. {\it Boundedness} : $|d(\alpha,\beta)| \leq 1$; this is suggested by the operational 
meaning  of $|d(\alpha,\beta)|$ as an intensity. Clearly the intensity of the final beam 
cannot be larger than the original one's. 
\\ \\
 7. {\it Subadditivity} : if $\alpha$ and $\beta$ are operationally additive, then  
$|d(\alpha,\alpha)| \leq |d(\alpha + \beta, \alpha + \beta)|$; from a coarser filter  the
 beam will exit  with  higher  intensity.

  \paragraph{Correlations}  As we mentioned, one can view the coherence functional as a 
functional on ${\cal A}_h \times  {\cal A}_h$. As such it  can be  defined on
 pairs of observables. In fact one can show that 
 \begin{equation}
 d(A_{t_1} \otimes \ldots \otimes A_{t_r} , A_{t'_1} \otimes \ldots \otimes A_{t'_n}) 
= G_A^{nr} (t_1, \ldots, t_r; t'_1, \ldots t'_r ) 
 \end{equation}
 where by $G^{rs}_A$ we denote the $rs$ correlation function, where $r$ denotes the  number of
 indices that are   time-ordered and $s$ the number of indices that are anti-time  ordered. 
The generator of these functions is known as the closed-time-path generating  functional 
associated to $A$ \cite{Schw61,Kel64}. As a functional on ${\cal A}_h$ it is identical to the 
coherence  functional. For our purposes, it is sufficient to remark, that the knowledge of 
 the coherence  functional allows us to fully reproduce any correlation function of the theory 
\cite{An00b}.

  \subsection{Conditioning} 
When we want to go beyond descriptions of ensembles and actually predict properties of an 
individual  system, we need to incorporate the information we obtained through experiments
 into the mathematical object  that allows us to make  predictions. This process is known as 
conditioning. In classical  probability theory it is implemented through conditional 
probability. In standard quantum theory, it is commonly known as "wave packet  reduction", 
even though in a strict sense it does not  correspond to a physical process. 
If a measurement corresponding to $P$ is realised then the state  of the systems transforms
 as $\rho \rightarrow P\rho P /Tr(\rho P)$.  

The consistent histories approach defines conditioning through probability theory. That is,
 in a consistent  set that classical probability holds, one can use its rules to define 
conditional probabilities. This choice comes from the insistence of the consistent histories
 approach on probabilities as the  basic objects of the theory.  From our perspective it is 
not necessary to use concepts of probability theory to define  conditioning. In fact,
 conditioning is primarily an algebraic rather than a probabilistic concept.  If we have a 
system described by the coherence functional $d$ and we have found that a filter  history 
$\alpha$ has been realised, then we condition as follows.

 First, we restrict to all filter  histories $\beta$ {\it operationally  compatible} with 
$\alpha$.    This means that if $\alpha$ is a series of filters, then any allowable setup 
has to include the filters of $\alpha$ or   filters that are compatible  and coarser from 
the  filters of $\alpha$. In this sense operational compatibilty means that there exists 
 a {\it filter history} $\gamma$ such that $\alpha \leq \gamma$ and $\beta \leq \gamma$, 
such that there is no other filter $\gamma'$ satisfying this property with
 $\gamma' \leq \gamma$. We shall denote such a filter history 
$\gamma = \alpha \circ \beta$ \footnote{In the language of lattices,  we demand that the meet
 of $\alpha$ and $\beta$ exists  is  a filter history itself.} .    Then we define the
 conditioned coherence functional  for pairs of filter histories 
 \begin{equation}
 d_{\alpha}(\beta,\beta') = d(\alpha \circ \beta, \alpha \circ  \beta')/ d(\alpha,\alpha)
 \end{equation} 

We divide over $d(\alpha,\alpha)$ in order to satisfy the normalisation condition.  
 For the simple case that $\alpha$ consists for a simple filter $P$ before all measurements 
 we have that $d_{\alpha}$ is given by the standard expression (3.2), with  the substitution
 $\rho \rightarrow P\rho P /Tr(\rho P)$. If $\alpha$ consists of a single filter $P$ at time
 $t_f$ after all measurements, the coherence  functional is given by 
 \begin{equation}  
d(\alpha,\beta) = Tr ( C_{\alpha}^{\dagger} \rho C_{\beta} \rho_f) \ Tr(\rho \rho_f) 
\end{equation}
 where $\rho_f = e^{iHt_f}Pe^{-iHt_f}/TrP$.  

 In our operational perspective,  conditioning means that we restrict our experiments, 
to having certain  filter configurations that correspond to $\alpha$  fixed. If one wished 
to make predictions about individual systems,  according to the consistent histories  interpretation, 
one can   use
 the conditioned coherence functional to define consistent sets and seek for histories with 
probability one in these sets. 

   \subsection{The relative phase theory} 
\subsubsection{Standard quantum theory as a model}
 The specification of an algebra  and of a class of coherence functionals constitutes a  model
 theory that aims to describe the filter experiments we described.   Quantum theory takes for
 ${\cal A}$ the set of bounded operators $B(H)$ on a complex Hilbert  space $H$. A filter 
then corresponds to a projection operator on  $H$. When we combine the  algebras to form the
 history algebra, we have  ${\cal A}_h  = \otimes_t B(H_t) =   B(\otimes_tH_t)$. This means
 that histories can be thought as (homogeneous) projectors  on  $\otimes_t H_t$, while general
 (time-averaged) observables can be identified with  self-adjoint operators on $\otimes_t H_t$.

 The algebra ${\cal A}$ is non-Abelian.  Two projectors do not generically commute, hence two 
filters need  not be compatible.  Clearly, when we study two systems we have $B(H_1) \otimes 
B(H_2) = B(H_1 \otimes H_2)$ and  the combined system is characterised by the Hilbert space
 $H_1 \otimes H_2$.  The coherence functional is given by equation (3.2). It is constructed out of 
two ingredients: one that  
  characterises the nature of the physical systems (the Hamiltonian)
 and one that characterises  the way the source $S$ was constructed. This second piece is 
represented by a   Hilbert space vector or more generally by a density matrix.   

\subsubsection{ Abandoning the Hilbert space} 
Quantum mechanics assumes that the algebra of filters is non-commutative, corresponding to  
the bounded operators in a complex Hilbert space. If one takes quantum theory to be primarily 
a theory  of probabilities, then non-commutativity as a requirement has a  deep physical 
meaning. The product of  two non-commuting self-adjoint operators is not self-adjoint, hence
 the correlation between them  is inevitably complex. The non-commutativity of two operators 
is  connected with  the need  of complex numbers in quantum theory, since both are necessary 
for the uncertainty principle  to hold \footnote{St\"uckelberg has showed that in a real
 Hilbert space, there is no uncertainty  relation between non-commuting operators, unless one
 introduces a complexification \cite{Stuc}.}.  

Non-commutativity is therefore an integral part
 of quantum theory, when this is viewed from the perspective of probabilities. But we
 have emphasised that quantum theory is {\it not} only  a theory of probabilities, rather 
it also includes information from relative phases. We have tried  to develop a chain of 
arguments leading to a scheme, where these phases enter in the formulation  of quantum theory 
as primitive ingredients: in the coherence functional.  We are then led to the natural question, 
whether non-commutative observables  and more generally the Hilbert space structure are 
{\it at all  necessary} in this scheme and whether  one can substitute them with a simpler 
structure without compromising the predictive power of quantum  theory.   

The answer to this last question is affirmative: one can do without  the Hilbert space structure. 
 In fact, we can get a theory that is based on commuting observables (simple functions),  
that  fully reproduces the predictions of quantum theory.    We  define the space of filters 
as in classical probability theory. They ought to  correspond to idempotent  elements of 
$B(\Omega)$, the set of bounded functions on  some (measurable) space $\Omega$. A
filter is, then, represented by a characteristic  function of a subset of $\Omega$. Clearly
 a history  corresponds to an element of $\otimes_t B(\Omega_t) = B(\times_t \Omega_t)$. 
Hence a history  corresponds to an element of $\times_t \Omega_t$, or rather a subset 
$\Omega_h$ of it  consisting of some suitable  maps from ${\bf R}$ to $\Omega$ .   
 Two subsystems are combined by $B(\Omega_1) \otimes B(\Omega_2) = B(\Omega_1 \times 
\Omega_2)$,  hence they  are represented in a space $\Omega_1 \times \Omega_2$.  And then
 we define a coherence functional as a  bilinear functional 
 \begin{equation} 
d: B(\Omega_h) \times B(\Omega_h) \rightarrow {\bf C} 
\end{equation}   
The essence of this construction  is that {\it the filters and the observables} are identical 
 to the ones of classical theory and the distinct quantum mechanical behaviour is encoded
 {\it  in the introduction of the coherence functional} (and hence the relative phases) as 
the primitive  elements of the theory. This class of theories we shall call 
{\it relative phase theories}. 

 Commutative variables are considered as fundamental in hidden 
variable theories, as for instance in Bohm's mechanics.  Another  instance,  is Nelson's
 stochastic mechanics \cite{Nel85}, which tries to reproduce quantum mechanics by a stochastic
 process on {\it configuration} space. Besides problems of locality, this construction 
cannot reproduce the unequal-time correlation functions of quantum theory. In fact,
 no additive probability measure can.  

Our proposal is closer in spirit and substantially influenced by  to Sorkin's quantum measure 
theory \cite{Sor94, Sor97}. We have a different attitude, though: we do  not agree  that
  measure-theoretical ideas are themselves sufficient to describe individual quantum systems.  
At the level of formalism, the main difference is that Sorkin  insists on the probability 
structure  (his quantum measure is the real part of the coherence functional), thereby 
downplaying the  importance of relative phases.  The reader is also referred to \cite{An97},
which states  different  motivations for this line of reasoning.

  \subsubsection{Construction of a theory} 
 In order to construct a relative phase
 theory for a physical system,  we would have to specify the space $\Omega$ of elementary 
alternatives and find a procedure that will  enable us to write a large class of physically 
relevant coherence functionals on this space.  The question then arises, whether there exist 
such constructions that reproduce the predictions  of quantum theory.

  In standard quantum 
theory it is known that one can obtain full information about a physical  system, solely
 through (unsharp) phase space measurements. The reason for this is that every physical
 system we know, has a phase space structure incorporated in its quantum description.
 This comes  from the canonical commutation relations, or stated differently, from the 
representation of the  canonical group \footnote{The canonical group is classically identified
 as a group acting transitively  by canonical transformations on a classical phase space. In
 the corresponding quantisation scheme  it is required, that the Hilbert space of the quantum 
theory ought to carry a unitary, irreducible  representation of the canonical group.
 Conversely, if a Hilbert space of a quantum theory admits  an irreducible, unitary 
representation $U(g)$ of some group $G$, we can construct the coherent  states 
$U(g) |0 \rangle$, where $|0 \rangle$ is a fiducial vector. If we define an equivalence 
 relation on $G$ such that $g \sim g' $, if $U(g)|0 \rangle $ and $U(g')|0 \rangle$, just 
differ  by a phase , the resulting space $G/\sim$ can be viewed as a phase space for the 
classical system. It carries a symplectic structure, which it inherits from the imaginary part
 of the Hilbert  space's  inner product. It also carries a metric structure coming from the
 real part of the inner product. }  on the system's Hilbert space. This suggests that the 
phase space ought to be  identified with the space $\Omega$ of a relative phase theory. 
And the representation of the canonical  group gives a way to construct coherence functional
 on phase space, that correspond to the  standard quantum mechanical ones. 
The procedure by which this is effected is known as the Wigner  transform.  

Systems that have a linear phase space have as canonical group the   Weyl group.  For the 
system case of  an one-dimensional system (particle at a line), its Lie algebra is  determined
 by 
\begin{eqnarray}
 [ \hat{q}, \hat{p} ] = i 
 \end{eqnarray} 

 Let us denote by $\hat{U}(\chi,\xi) = e^{i\hat{q} \xi + i \hat{p} \chi}$ the unitary operator
  representing one element of this group. Then we can define the operator 
 \begin{equation} 
\hat{\Delta}(q,p) = \int d \chi d \xi e^{ -i q \xi - i p \xi} \hat{U}(\chi,\xi) 
\end{equation} 
The operators $\hat{\Delta}$ provides a linear map from the space of operators on the  
Hilbert space $H$ to the phase space $\Gamma$ of our system 
 \begin{equation} 
\hat{A} \rightarrow F_A(q,p) = Tr(\hat{A} \hat{\Delta}(q,p)) 
\end{equation}
 This map is trace preserving, in the sense that 
 \begin{eqnarray} 
Tr\hat{A} = \int \frac{dq dp}{2 \pi} F_A(q,p) \\ Tr ( \hat{A} \hat{B})
 = \int \frac{dq dp}{2 \pi} F_A(q,p) F_B(q,p) 
\end{eqnarray}
 But it does not preserve the multiplication; the condition $P^2 = P$ is not preserved  in 
phase space: a projection operator is {\it not} mapped into a characteristic function. A 
sharp quantum mechanical filter is not sharp on phase space: this is a manifestation of the
 uncertainty principle.  Since the Hilbert space,in which histories live is constructed out 
of tensor products  of the single-time Hilbert space, the Wigner transform can be employed to
 pass from the  Hilbert space $\otimes_t H_t$ to the space of histories. Indeed, for any
  operator on $\otimes_i H_{t_i}$, $ i = 1, \ldots, n$, we define a function $F_A$ on
 $\times_i \Gamma_{t_i}$ as 
\begin{equation} 
F_A(q_1,p_1,t_1;\ldots; q_n,p_n,t_n) =
 Tr_{\otimes_i H_{t_i}} \left( A (\hat{\Delta}(q_1,p_1) \otimes\ldots \
 \otimes \hat{\Delta}(q_n,p_n) \right) 
\end{equation} 
  Given any discrete-time history with support $\{ t_1, \ldots, t_r \}$ we can define 
 the operator $C_m$ as
 \begin{eqnarray}
 \hat{C}_m = e^{i\hat{H}t_1} \hat{\Delta}(q_1,p_1) e^{-i\hat{H}t_1} \ldots e^{i\hat{H}t_m} 
 \hat{\Delta}(q_m,p_m) e^{-i\hat{H}t_m}  
\end{eqnarray}
 and define the following object that has support on a pair of an $n$-point and an 
 $m$-time temporal support.
 \begin{eqnarray}
 W_{n,m}[q_1,p_1,t_1; \ldots q_n,p_n,t_n| q'_1,p'_1,t'_1;\ldots ; q'_m,p'_m,t'_m] = 
 Tr \left( \hat{C}_n^{\dagger} \hat{\rho}_0 \hat{C'}_m \right), 
\end{eqnarray}  
 Then it corresponds to a coherence  functional on 
$\otimes_i B(\Omega_{t_i}) \times \otimes_i B(\Omega_{t'_i})$.  To a pair of functions 
$A \in \otimes_i B(\Omega_{t_i})$ and $B \in  \otimes_i B(\Omega_{t'_i})$  it assigns the complex 
number  
\begin{eqnarray}
 d(A,B) = \int \frac{dq_1 dp_1}{2 \pi} \ldots \frac{dq_n dp_n}{2 \pi}  
\frac{dq'_1 dp'_1}{2 \pi} \ldots \frac{dq'_m dp'_m}{2 \pi} \nonumber \\
  A(q_1,p_1,t_1;  \ldots q_n,p_n,t_n) B(q'_1,p'_1,t'_1, \ldots, q'_m,p'_m,t'_m) \nonumber \\
 \times W_{n,m}[q_1,p_1,t_1; \ldots q_n,p_n,t_n| q'_1,p'_1,t'_1;\ldots ; q'_m,p'_m,t'_m] 
\end{eqnarray}    
 In fact, one can prove that these discrete time definitions allow for a definition of the 
 coherence functional for continuous-time histories \cite{An00b}. There are two structures 
on phase space, that are relevant  in this case: a $U(1)$ connection form $p dq$ that is 
responsible for the introduction  of the complex numbers in the coherence functional and the
 Moyal bracket. The later is   represents the algebra of  operator commutators on phase space.
 The Moyal bracket $\{\{ ,\}\}$ between a pair of functions on phase space $f$ and $g$
 reads
 \begin{equation} 
\{\{f,g\}\} = 2 i f \sin \left( \frac{1}{2} \{,\} \right) g , 
\end{equation} 
where by $\{,\}$ we denote the Poisson bracket as a bilinear operator: $f \{,\}g = \{f,g \}$. 
 The theory, thus constructed gives the same values for the correlation function of any 
 observable.  Any quantum mechanical operator corresponds to a function on $\Omega$. 
Its correlation  functions are   identical with the quantum mechanical  ones, by virtue of 
equation (3.16).    But in the relative phase theory the sharp filters are different from the
 corresponding  quantum mechanical ones. Hence the predictions about outcomes of idealised, 
 "precise" measurements  would differ and so would the corresponding relative phases.  Note, 
that even at a single moment of time, the theory is not described by a probability 
 distribution. This is equivalent to the well known fact, that the Wigner function  is in
 general non-positive. In terms of filter measurements, this means that the relative phase 
 between two single-time histories with incompatible filters is generally non-zero. In 
standard  quantum theory (and for sharp filters) this vanishes identically.  

 This construction preserves naturally all properties of the coherence functional, since it 
amounts  to substituting in equation (3.2) general positive operators in place of projectors. 
In particular, property 6. also holds since a characteristic function corresponds to a 
positive operator with  norm less than unity, which is a sufficient for a proof.  

  The Wigner transform can be generalised for other quantum mechanical systems. For spin systems,
it arises from the study of the representations of SU(2)
 \footnote{The phase space for a single fermionic  oscillator out of which the 
field is constructed is the two sphere $S^2$.}, while for bosonic fields one uses the infinite-dimensional 
Weyl group as the canonical group. Even for fermion fields there exists a phase space
 description \cite{Kla59}.
  Hence, for all systems of physical interest 
one can construct a relative phase theory, that completely  reproduces the predictions 
of quantum mechanics.  

 We have to add here, that the Wigner transform is not the only possible group-theoretic
 construction, that  can give predictions that reproduce the ones of quantum theory.
 There exist also the $P$ and  $Q$ transformations that are based on coherent states. 
The Q-transform of an operator $A$ is its expectation value on the coherent state basis 
$Q_A(x,\xi) = \langle x \xi|A|x \xi \rangle$,  while its P-transform is a function $P_A(x,\xi)$
 defined by $A = \int dx d \xi P_A(x,\xi) | x \xi \rangle \langle x \xi |$. The following 
 property holds \begin{equation} Tr(AB) = \int dx d\xi Q_A(x,\xi) P_B(x,\xi) = 
\int dx d \xi Q_B(x,\xi) P_A(x,x\i)
 \end{equation} 
This implies, that we could construct  then a relative phase theory by considering either
 the P-transform of the coherence functional together  with the Q-transform of operators for 
the filter algebra, or conversely the Q-transform for the  coherence functional and the $P$ 
transform of the operators of the filter algebra. And perhaps this does  not exhaust the 
possible ways of constructing a relative phase theory on phase space, that  reproduces the
 correlation functions of quantum theory. Which one  of them is the correct  description is 
something that quantum theory itself cannot answer. We would need a theory  for the 
individual system.     

 \section{General properties}  
 To summarise: in the previous sections we argued that, 
at least at an operational level, quantum  phenomena can be described in terms of a filter 
algebra ${\cal A}$ and a coherence functional that  gives the relative phase and probability
 content of this theory. And then we proposed, that \\ \\ 
i. Quantum theory can be described by a commutative filter algebra, corresponding to functions
  on some space $\Omega$.
 \\ \\
 ii. The space $\Omega$ can be identified with the phase space of the corresponding classical
  theory.   \\ \\
 In general we can use group theoretic constructions (like the Wigner transform)  to 
construct a coherence functional on $\Omega$, that reproduces the predictions of  quantum 
theory.  We are now going to analyse the distinct structures and general  properties that our 
proposal implies. 
 \subsection{Quantum logic}
 We said earlier that the space of filters has a partial ordering relation. We also said that 
in special  cases it is operationally meaningful to consider the addition as corresponding to
 a conjunction of propositions. In standard quantum theory, filters correspond to projection 
operators on a Hilbert  space, and it turns out that the partial ordering structure is complete
 and forms a {\it lattice}.  This lattice has a number of operations that are algebraically
 identical to the conjunction ($\bigwedge$), disjunction ($\bigvee$) and negation of 
logical propositions. It is therefore often said that the lattice of projectors  on the Hilbert
 space is the logic of the quantum theory, which is distinct from classical  logic. Indeed, it
 is stated, that a projection operator corresponds to a proposition about  a property of an  
 individual physical system. This is motivated by the analogue in classical probability, that a
  statement about a physical system can always be phrased as a statement that the system lies 
in  a subset of the sample space $\Omega$. And the set of subsets of $\Omega$ has the structure
 of a Boolean  lattice. 

   The interpretation of the algebraic structure of a lattice as the 
logic of individual systems  is  often questioned,   due to the fact that the lattice of
 projectors is not distributive. This means that  if $\bigvee$ and $\bigwedge$ represent
 the algebraic operations of disjunction and conjunction between projectors it is not always 
true that
  \begin{equation}
 P \bigwedge (Q \bigvee R) = (P \bigvee Q) \bigwedge (P \bigvee R)
 \end{equation}  
The failure of distributivity is the cause of the {\it Kochen-Specker theorem}: one cannot 
 consistently assign  true or false values to all propositions. 
In other words the lattice
  of projectors cannot be taken to describe properties of an individual system 
at a single moment of time.
 Properties  of
 such systems are at most contextual. We referred earlier  a manifestation of this  pathology 
in  the  consistent histories approach. If contextuality is viewed as referring to our 
perspective of the system, the notion of objectivity is lost. If the context, it refers to, is
 a concrete  experimental setup, there is little gain  from the Kopenhagen interpretation.
 For instance, in consistent histories there are very few propositions, beside  measurement
 outcomes that can be considered as ``true'' for a physical system \cite{DoKe96}.    

We have stressed throughout, that we deliberately take an  operational stance. In fact, we are
  rather wary to consider, that the models for beam measurements  have validity as description
   for individual systems,  and even  more wary to talk about logic of such systems.
 However, unlike the standard quantum theory, {\it should we  wish to do so, we can}. 
We are not constrained by the Kochen-Specker theorem. Our filters correspond to  subsets
 of $\Omega$, hence to a distributive lattice. The "logical " structure of our theory is 
identical  to the one of classical probability and therefore completely unambiguous, as far
 as definability of properties is concerned \footnote{Note, that there is no reason to assume 
 that  our logic is Boolean. A Boolean lattice is necessary in probability theory,  for it is 
needed in  the definition of measures. Our relative phase construction is not a probability 
theory. It could be that it has a different structure as fundamental, e.g. continuity. But 
in any case, our "logic" is governed by the laws of intersection and union of sets and 
therefore will always be  distributive. This is sufficient to remove the spectre of the 
Kochen - Specker theorem.}.  

\subsection{Predictability}  
Usually predictability refers  to our ability to make predictions about an individual system,
 when we  have some probabilistic description of it. In classical probability, one says that 
if the probability  of an event is $1$, then almost surely this event will be  realised for 
any individual system.  In quantum theory predictability is a more complicated issue; the
 Kopenhagen interpretation  does not care to address it: an operational treatment needs not be 
concerned with individuals. And with good reason:  prediction and retrodiction are obscure in 
all realist schemes. Amongst such schemes, the  consistent histories provides the most solid
 treatment. It  states, that if in a consistent  set the probability of an history is equal to
 one (or if the conditional probability of an event is one and the condition is satisfied),
 then this event is predicted ( or retrodicted) by the theory.  Unfortunately, (at least) 
retrodiction is pathological. In different consistent sets one can retrodict mutually
  exclusive propositions \cite{Kent97}. This is again the problem of contextuality.  

  Now, in the relative phase theory we can have a consistent histories formulation  of 
predictability (after all the formalism satisfies the Gell-Mann -  Hartle - Isham axioms 
\cite{Har93a,I94}). Let us take two histories $\alpha$ and $\beta$ that are disjoint
 and exhaustive ($\alpha+ \beta = 1$). Consider that  $d(\alpha, \alpha) = 1$. Would it mean 
that we can definitely predict that $\alpha$ will  be realised \footnote{The coherence
 functional could have been obtained by the incorporation of a  number of experiment outcomes,\
 using conditioning as described in section 3.4.} ? Since we have $d(\alpha+\beta,\alpha + 
\beta) = 1 = d(\alpha,\alpha)$, we would have  that $d(\beta,\beta) = - 2 Re d(\alpha,\beta)$.
 Hence it is not as though the intensity of the  beam passing through $\beta$ vanishes. 
The possibility $\beta$ cannot be ruled out unless $Re d(\alpha, \beta) = 0$. This is the
 consistency condition. In this case, our chosen set of histories is described by  a 
probability measure. Then one can use a rule of inference that states: whenever a probability 
 for an event is unity, then this event is predicted (or retrodicted) within our consistent
 set.  The dependence of predictions upon choice of sets would carry out here as well.

 We would then need to check whether there are incompatible predictions in different 
consistent sets. We will examine  
the most elementary  case:  we take a coherence functional, perhaps conditioned with respect 
to some experimental data. Then consider   three disjoint alternatives $\alpha,\beta,\gamma$, 
that are exhaustive ($\alpha+ \beta + \gamma = 1$). Let us assume that 
$\{ \alpha, \beta + \gamma \} $ forms a consistent set in which $\alpha$ is predicted. 
This means that $d(\alpha,\alpha) = 1, d(\beta,\beta) = 0 , Re d(\alpha, \beta) = 0 $. 
And let us also assume that $\{ \alpha + \beta, \gamma \}$ forms a consistent set for 
which $\gamma$  is predicted. This is clearly a situation, where we get incompatible 
inferences in different  consistent sets. We can verify,  that  for this to occur we need
 have  $Re d(\alpha,\gamma) \leq - \frac{1}{2}$.   This means that the contextuality of
 propositions is not necessarily  a consequence of  non-distributivity. In our case, 
it would come from the non-additivity of the probability measure, together with the   
prediction rule. 

Of course, since we have a distributive logic,  the incompatible inferences are much more 
controllable.  They cannot arise if $Re d(\alpha, \beta) \geq - \frac{1}{2}$. One could be
 tempted to impose this restriction on allowable physical theories, or use it to  define 
{\em strongly predictable} quantum theories. 
  But we believe, that the notion of prediction based 
on consistent sets is rather artificial  for the nature of quantum theory. Fundamentally, 
the relative phase theory is {\it not} a  probability  theory  and an attempt to force it into
 the strict axiomatic framework  of Kolmogorov probability,   does violence to  its nature.
 We are therefore very skeptical, whether this formalism can be extended  to describe
 properties of an individual system and make predictions about it.  We are content with the
 description of beam experiments  and anything that can be modeled upon them. Our attitude
 is very much  into the tradition initiated by Bohr; but unlike him we have no reason to 
believe that our formalism  is complete. We have defined it for measurements of ensembles 
and it is for measurements of ensembles  only that it is good.  

  The most we could say about
 the decoherence condition, is that there is a degree of coarse-graining  in the filters, 
that would allow us to ignore all phase information and describe the filter  experiments using
 another model: classical probability theory. And within this approximation, one  could
 sometimes  interpret the theory in such a way as to  make some predictions about an
 individual system. But probability theory arises as {\it an approximation}, not as a
 fundamental set of concepts that {\it interpret our theory}. 

  \subsubsection{The classical limit}
 The observables of our theory are functions on a space $\Omega$, which we can take to be 
 the classical phase space of the system. When sufficient coarse-graining is allowed as to 
 make the  approximation by probability theory sufficiently good, the system will be described
 by  a stochastic process on phase space. Typically, for sufficient coarse-graining, this 
process is  almost deterministic giving rise to the Hamilton equations of motion. 
  
There is no ambiguity what the classical limit of the theory will be and  that it is 
 independent of the degree of coarse-graining. This is unlike  the consistent  histories 
approach, where the non-distributivity of the lattice of propositions  makes in principle 
possible the existence of very different and incompatible classical limits (what Gell-Mann and
 Hartle call quasiclassical  domains). In fact,  such  quasiclassical 
domains are generally unstable \cite{Kent96}, something  that diminishes predictability even\
 at the semiclassical level. 

  To summarise: we do not believe that there is yet  a meaningful way to make predictions
 about individual  systems. Only in the case, that we consider sufficient coarse-graining, 
 so that the decoherence condition  approximately holds, we can approximate our theory by 
classical probability theory. This, together with a set of assumptions that have to do 
with the meaning of probabilities in a classical setting,    might allow us to make 
predictions about an individual system. But  it is by means of an approximation that we can
 make predictions,  not by means of a fundamental law of nature.  

\subsection{The Bell-Wigner theorem}
 Any new interpretation of quantum theory that tries to do without the Hilbert space has to
 phase  the restraining demands of Bell's theorem and all its generalisations.   Bell's
 theorem is usually taken to imply that realism and locality is not compatible with quantum 
theory. By realism, one usually means the specification of hidden variables, i.e. variables
 that characterise  more precisely the state of the system. 

Our recipe, that the underlying 
structure 
 of quantum theory  is the classical phase space implies  the use of the Cartesian product to 
combine the
 phase space of such systems. This  might lead to   a hasty conclusion that such a construction is
 forbidden by Bell's theorem.   This is not true. The most general proof of Bell's theorem is 
based on the assumption that the ensemble description of   hidden variables  that characterise 
individual systems, {\it is given  by a probability distribution that satisfies the Kolmogorov 
axioms}. This is exactly what we explicitly renounce in this paper. We do not believe that
 probability theory is   {\it a priori} of relevance to a physical theory. Probability theory 
is a branch of mathematics  that has provided useful models for certain physical phenomena 
and there is no reason to expect  that it would be relevant for the totality of 
them \footnote{Probability theory is not as  primitive   a structure as arithmetic  or
 geometry is.  These can be argued to underlie most, if not all, of our physical concepts 
and form an irreducible mathematical background from which to view physical phenomena.  
Unlike them, probability theory arose much later and its relation to physics was not  
immediately  evident. People became confident with its use in physics after the  success 
of statistical mechanics, and  this description was put in a solid mathematical footing by 
Kolmogorov. And still, the use of  probability theory   had to be justified by physical 
arguments: the ergodic postulate.  We should also add, that Kolmogorov probability, which is
 a measure-theoretical,  description of probabilities is but one formalisation of the
 intuitive notion of probability. Its abstractness and the  rich  mathematical structure 
of measure theory, make it extremely useful, but its applicability to a particular  
situation should be  a matter of the {\it physics} of the concrete physical system. It 
is easy to think of counterexamples in classical physics, of ensembles with random behaviour,
 that {\it cannot be described by a probability measure} \cite{Khre00} }.

 Quantum theory also shows that  it is not only frequencies of occurences that we measure, but
 also relative phases, which do not  fit any concepts of probability theory. Once we remove 
this condition, Bell's theorem  is not constraining.  In this regard let us note the
 following points. Bell's original proof did not use a  probability distribution. He assumed 
 definite values for each particle of the  correlated pair, which essentially implied 
 {\it determinism}.  As such, it is not constraining for our formulation. There exist more 
general versions  of Bell's theorem that employ probability distributions   and  constrain 
stochastic hidden variable theories. They are not relevant to our case.  There is also the 
Greenberger-Holt-Zeilinger (GHZ) argument \cite{GHZ}, which pertains to refer to properties
 of individual  systems and distinguishes the prediction of hidden variable theories from 
the outcome  of a single measurement. In their proof they use an assumption from probability 
 theory: if the  ``probability'' of an event is equal to $1$ then this will be true. As we
 showed in  the previous section, this is valid  only if the probability satisfies the
 Kolmogorov  additivity  condition. As such it cannot be used against a hidden variable 
theory  that is not modeled this way.  

Overall, we believe that we should make the qualification that:  Bell's theorem forbids not
 local realism in general, but local realist theories that  {\it are modeled by classical 
 probability of the Kolmogorov type}. In the  light of our discussion, it is a much weaker 
restriction and has fewer metaphysical  implications,   than what is usually claimed.  
 
We should note that there exists a group theoretic justification for the use of Cartesian
 product  to construct the phase space of the combined system.
 If $\Gamma_1$ and $\Gamma_2$ are phase spaces and $G_1$, and $G_2$  the corresponding 
canonical groups, then $G_1 \times G_2$ is the canonical group of  $\Gamma_1 \times \Gamma_2$.
 If $G_1$ is irreducibly represented on a Hilbert space $H_1$ and $H_2$ on a Hilbert  
space $H_2$, then $G_1 \times G_2$ is irreducibly represented on $H_1 \otimes H_2$ and 
a Wigner transform can  be naturally constructed to pass from the quantum theory on
 $H_1 \otimes H_2$ to the relative phase  theory on $\Gamma_1 \times \Gamma_2$. 
This construction allows us to fully reproduce the  correlation functions of the quantum 
theory: in fact, it was done by Agarwal in reference \cite{Aga93}. 

 \subsection{The uncertainty principle} 
The Wigner transform in standard quantum theory yields a non-positive function  on phase
 space (the Wigner function) corresponding to  a density matrix. It does not define a 
probability distribution. This is usually taken to  be a consequence of the noncommutativity
 of position and momentum  and of the fact  that one cannot measure position and momentum with
 infinite accuracy. These remarks are correct,  when they refer  to beams, i.e. ensembles.
 But it is often stated that there is no meaning to a  sharp phase space point for an 
individual system. This  is completely unwarranted by the precise  formulation of the 
uncertainty principle: $\Delta q$ and $\Delta p$  are defined  as deviations for  a series 
of measurements of position and momentum respectively in an ensemble of physical systems 
\footnote{The operator form of the uncertainty principle has a different meaning from the
 original derivation of Heisenberg. Heisenberg discussed the physical mechanisms by which
 measurements of a single physical system are  prevented from giving a simultaneous accurate
 reading of position and momentum. But neither Heisenberg's derivation implies
 the non-definability of sharp phase space properties. In standard quantum theory 
this is only implied by the Kochen-Specker theorem.}.

   In our scheme the uncertainty principle still holds, since it refers to  the comparison of 
 results of measurements with  filters, that correspond to  phase space observables. Both
 $\Delta q$ and $\Delta p$ can be measured as correlation  functions by a sequence of series 
experiments.  However, in principle,  we can have   filters that correspond to a phase space
 area  less than $\hbar$. The beam passing through them,  will develop  a much stronger phase 
shift, compared to the case, where  it  passes from a coarser  filter. In effect, if the 
Wigner function of the beam oscillates around zero, taking negative values, at some phase 
space scale, it is  filters with accuracy in this scale, that will give rise  to  a strong
 interference behaviour for the beam. Nonetheless, both intensities  and relative phases are
 well predicted by the theory.  

 \subsection{Elements of reality}
 We have  refrained from considering our formulation as anything more,  but an operational one,
 in the spirit of Kopenhagen.  We can, then, have no claim about existence or not of 
 elements of reality for an individual quantum system.  

 However, if this were out true aim, 
we would not have bothered to move beyond   Hilbert space quantum theory. The main reason for
 abandoning Hilbert space for the phase  space is that the latter is more  amenable to our
 intuition for individual systems and has a rich geometric structure.  We would like  
 to formulate a geometrical theory on phase space for individual systems,  that will explain 
 the role of the coherence functional and its relation to statistics of ensembles. In a sense, we want 
to ask the following question: what geometric behaviour on the phase space leads to a statistical behaviour 
like the one given by a coherence functional? In a sense, this question is 
analogous to the one about  the validity of the description of macroscopic systems by statistical mechanics and 
probabilistic concepts. There, an answer, was the property of ergodicity or quasiergodicity. Would it be possible to 
find a similar answer in the quantum case?

 The only thing we can say right now is 
that  the coherence functional  is 
built from geometric objects on the phase space: bundles and connections \cite{AnSa00,An00b}.
These  structures arise naturally   
 in the  geometric quantisation scheme \cite{Wood} and the group 
theoretic approach to quantisation \cite{I83}. The temporal logic
 histories programme \cite{I94, IL95, ILSS98} also suggests that these geometric structures 
are related to the temporal   structure of this theory \cite{Sav99a, Sav99b}. These are 
 interesting  links that will  set the tone of the work to follow.

What the present      construction has accomplished  is   the removal of the constraints of Bell
 and Kochen-Specker's theorems. This  makes plausible the existence of  a geometric 
description for individual systems that  reproduces the ensemble predictions of Kopenhagen
 quantum theory. In this respect, our  stance is very much a continuation of  Bohm's  
programme. But all our motivation and  structural insight has come from the consistent
 histories approach to quantum theory.

    \subsection{Experiments} 
The only difference between our scheme and standard quantum theory, lies in the specification 
 of the sharp filters. In an ideal world, where all filters could be assumed perfect, we would 
be able to explicitly  distinguish between the predictions of the relative phase theory and
 standard quantum theory. 

Unfortunately, this is not true. Even in standard quantum theory, a realistic filter is  
 best described by a positive operator (sometimes known as an effect). On the other hand, it
 is the construction of the filter that determines, what exactly it will let pass and one would
  need to have a detailed specification of its physics, before estimating what the function 
 that characterises it would be.    For these reasons, it is very difficult to imagine 
realistic experiments, that would be  able to distinguish  between these theories. 
Such an experiment would be of high  importance, as it could be said to separate between 
classical and quantum logic.   

 In general, the difference between the Weyl transform of a projection  operator and a 
characteristic functions on phase space is to be found in a phase space region  of the
 size of $\hbar$. This means that the distinguishability between classical and quantum sharp 
filters  is particularly accute in particle systems. There, the phase space is non-compact, 
hence the difference  of  filters at the order of $\hbar$ is practically impossible to detect.

    However, for spin systems the phase space is a two-sphere and its volume with respect to 
 the natural metric induced by the representation of $SU(2)$  (see footnotes 10 and  11) is 
of the  order of $\hbar$. In this case, one might expect a significant deviation in the
 measurement  of intensities or relative phases. Of course, this deviation would have to 
correspond to  the same correlation functions for the observables.   There is an important
 difference between classical filters and projection operators, which  comes from the fact
 that the Wigner transform does not preserve positivity. The Wigner  transform of a projection
 operator is a non-positive function, while a classical filter, even  if not sharp, 
cannot help not be positive. The negative values for the symbol of the projector  are
 particularly distinctive in the  case of spin systems. In this case, we have the symbol 
 corresponding to a projection on a vector characterised by spin in the ${\bf J}$ direction 
\begin{equation} 
F_P(\theta,\phi) = \frac{1}{4 \pi} ( 1 + 2 \sqrt{3} {\bf \hat{n}}\cdot {\bf J} ),
 \end{equation} 
where ${\bf \hat{n}}$ is the unit vector in the direction specified by $\theta$ and $\phi$.
 We see that it takes negative values in a significant portion of the sphere. Our initial  
thought  for a distinguishing feature was to exploit this lack of positivity in order to
 establish  a bound between the predictions of quantum theory and relative phase theory for 
{\it the statistical  correlation functions of the observables}. Two theories that have the same
 quantum mechanical  correlation functl  correlation functions do not have the same statistical ones, because in 
the latter there is  a different combination of  probabilities for elementary outcomes.  

Unfortunately, there are two flaws in this line of reasoning. First, it is not absolutely  
necessary that the transforms of the quantum mechanical projectors take  negative values.  
We could have  chosen to define the relative phase theory with respect to the Q-transform for 
the observables  and the P for the coherence functional. The Q-transform of a projector is 
always positive,  so the distinguishing feature between classical and quantum filters would 
not appear.  So, these distinctions would be model-dependent, rather than relevant to the 
basic structure  of a relative phase theory.

  But there is a second reason, which we believe is potentially more important, as it might 
 provide the beginning of an explanation, why the Hilbert space description is so natural 
 in quantum theory. Any physical filter is made out of matter and has to interact with the 
measured  physical system. One can, to first approximation, ignore all backreaction effects.
  But even  if the self-dynamics of the measured system is zero, there will be always be
 a non-zero  Hamiltonian evolution, due to the coupling with the degrees of freedom of the 
filter (in the case of spin consider the use of a Stern-Gerlach device as part of a filter).  
 The point is that (at least in the Wigner  picture) {\it the classical filters are not
 robust under quantum mechanical evolution}.  This is to say, that evolution as given by 
the Moyal bracket does not preserve positivity  of the filters. In this sense, a classical 
filter is an approximation to a realistic  filter. It is an idealisation for a non-material 
way of blocking   the beams. This idealisation is in essence  a fundamental part in a 
mathematical formalisation of the notion of experiment, as carried out in Kolmogorov 
probability theory.

 An analogy with 
classical probability might be indicative of what we have in mind. In analogy with the Heisenberg picure 
one can consider that the probability distribution of the beam stays the same, but the 
observables - filters evolve
in time. Hence, one has a differential equation $ \frac{\partial}{\partial t} P = {\cal L}P$, where 
${\cal L}$ is a differential operator, that incorporated the dynamics; in a measurement 
situation {\em it would contain a term for the interaction  between the beam and the device}. 
Now, in the case that all eigenvalues of
${\cal L}$ are {\it negative} and if we assume that the process lasts long, the smallest (in absolute 
magnitude) eigenvalue dominates and $P$  goes to its corresponding eigenfunction. Hence,  
the effective filter associated with a measurement is a function that solely depends upon 
the dynamics. It will not be negative-valued though, because physical evolution operators 
preserve positivity.
 Different measurement devices correspond to different forms of ${\cal L}$ and  different 
eigenfunctions.

This description is meaningful  also in the quantum case. The (Heisenberg type) equations of motion 
on the phase space $\Gamma$ read for an observable $A$ 
 \begin{equation}
 \dot{A} = \{\{A,H\}\} := {\cal L}_H A
 \end{equation}
 where $H$ is the Wigner
 transform of the Hamiltonian   and $L_H$ is an operator  on the space of phase space 
functions. We repeat that in a measurement situation $H$ ought to include the action 
of the classical device on the quantum system. 

 The space of phase space functions can be made into a Hilbert space $L^2(\Gamma, d \mu)$, through  the introduction of  the
 natural measure on $\Gamma$. Hence $L_H$ is an operators on  $L^2(\Gamma,d \mu)$.  
The classical analysis would be also valid in this case. If ${\cal L}_H$ has only 
non-positive eigenvalues, in the long time limit, any observable  would converge to an 
eigenfunction of ${\cal L}_H$, often a zero eigenstate. In the canonical Hilbert space picture 
this would imply that the effective filter, would be described by an operator commuting with 
the Hamiltonian $H$.  In the minimal coarse-graining case, this would be a projector 
 onto an  eigenstate of the Hamiltonian. Hence, if we assume that the time-scale of a measurement 
is much larger than the natural time scale associated to the Hamiltonian, it is natural mathematically 
to consider that the effective filter, would correspond to an eigenstate of the (interaction) Hamiltonian.

 In light of these considerations, it is possible, that the role of quantum mechanical filters 
is more  important in realistic measurement situations, {\it because of the nature of the averaged 
dynamics in the ensemble}. This would imply, that the underlying ``logic'' of the theory can be
 distributive, but  in ensemble measurements, projectors on a Hilbert space might  provide a more
 realistic description, when the interaction {\it dynamics} and the
 finite-time interval of a measurement process is taken into account.

 But the argumentation, we have presented,  is far from complete and at best only suggestive. 
The fact is,  we do not yet have a complete picture of how the standard Hilbert space would 
naturally arise  from the theory on phase space. However, this discussion suggests that
 the use of Hilbert space vectors might be a consequence not  only of the details of an 
underlying theory, but also from the basic operations one need  to perorm before setting 
an experiment. This would be something very desirable, since it  would be a justification of 
the fact that Kopenhagen quantum theory is the most efficient  way to describe statistical 
outcomes of ensembles. 

  In any case, these two arguments make unlikely, that it is possible
 to differentiate  between the statistical predictions of quantum theory and a relative phase 
theory.  It is  not impossible in principle, 
but it is difficult  to establish  a compelling argument
 that would  sharply distinguish between those two. Perhaps, we shall be able to devise one,
 when  we construct  a more definite theory, about the quantum behaviour of individual systems 
on phase space.  The few specifications we have given for a relative phase theory here, do not
 constrain much in  a degree sufficient to phrase definite statements, about predictions 
distinct from quantum  theoretical ones.  

  \section{Appraisal} 
  Let us first summarise the arguments, that are  central to the thesis of this paper 
\\  \\
   1. There is no {\it a priori} reason to assume that an ensemble of physical systems, is 
 describable by a probabilistic model that satisfies Kolmogorov's axioms. One has to 
 give a physical reason, in order to justify this assumption.
 \\ \\
 2. Quantum theory is not a theory of probabilities only: it also predicts relative  phases
 between different histories. These phases are measurable.
 \\ \\
 3. Any theory purporting to describe quantum phenomena needs to specify an algebraic 
 structure for the observables and a bilinear coherence functional that contains
  probability and phase information.
 \\ \\
 4. Once we accept phases as primitive ingredients of the theory, there is no  compelling
 physical  reason to employ  the Hilbert space formalism. Observables can be defined in a 
 purely classical fashion, as functions on some space of elementary alternatives.
 \\ \\
 5. The most conservative approach is to take for observables  functions  on
 the classical phase space. The Wigner transform is one way that  can  be used to construct a  coherence 
functional that fully reproduces the predictions of standard quantum theory. 
\\ \\ 
6. The resulting theory can be called a hidden variables theory, but does not suffer  
from Bell's theorem, because it is not a probability theory. It neither suffers from  
the Kochen-Specker theorem, because its logic is distributive. 
\\ \\ 
7. This construction suggests, that one should look on a geometric superstructure on  the 
classical phase space, in order to construct a viable quantum theory for individual  systems. 
\\ \\  \\ 
 A significant part in our argumentation has been the identification of an axiomatic 
 framework distinct from classical probability, that would  also be operationally  meaningful.
 This framework, we claimed, could substitute Hilbert space quantum theory; we,  therefore,
 focused   on the  mathematical possibility  of  reproducing   quantum mechanical predictions 
by a different model theory. For this reason  our constructions were mainly technical and 
no physical principles were evoked. This is what we need in order  
to establish  a  framework, that would allow us to  derive these results
 without any reference to quantum theory.

The second point is that there is no easy way to recover the  Hilbert space of the standard 
theory. We would expect a description in terms of Hilbert  space vectors  and the
 Schr\"odinger equation to arise in a simple manner and  to be related with the operational
 procedure of measurements.  We have not been able to find an intuitively simple way to do so. 
It seems  that the properties 1-7 for the coherence functional provide little restriction or  
insufficient guidelines for this purpose. An improved theory should have additional 
 assumptions, but these would only come from a detailed construction of a theory  for
 the individual system. Axioms  of statistical nature are, perhaps,  not sufficient by  themselves to
 explain  the structure of standard quantum theory.

   In any case, we do not purport to have  
a definite theory yet. This work  is more an indication of possibilities, rather than a 
completed framework.   We showed, that the Hilbert space formulation and concepts are not 
necessary in a  formulation of quantum theory, as far as statistical properties of
 measurements are  concerned.  This  removes  any impossibility objections and makes plausible,
   that the relative phase theories we described, are the  statistical limit of  a 
geometric theory of individual systems. Our main motive is the hope for a realist formulation 
of quantum  theory, which would tell us more about elements of reality than Kopenhagen does. 

\section*{Acknowledgements}
I would like to thank N. Savvidou for many discussions on the nature of quantum theory.
 I have also greatle benefited from discussions with Chris Isham, Jim Hartle, Adrian Kent,
 Rafael Sorkin, Bei Lok Hu and Greg Stephens. 

The research was supported by the NSF grant PHY98-00967.


\begin{thebibliography}{}  
\bibitem{vNeu} J. von Neumann,
 \newblock {\it The Mathematical Foundations of Quantum Mechanics.} 
\newblock (Princeton University Press, Princeton, 1996).  

\bibitem{Eve} H. Everett,
 \newblock " Relative State Formulation of Quantum Mechanics",
 \newblock {\it Rev. Mod. Phys.} 29, 454, 1957. 

 \bibitem{Gra73} B. DeWitt and N. Graham (eds.),
 \newblock {\it The Many Worlds Interpretation of Quantum Mechanics}, 
\newblock (Princeton University Press, Princeton, 1973).  

 \bibitem {Gri84} R. B. Griffiths.
   \newblock " Consistent Histories and the  Interpretation 
of Quantum Mechanics ", 
\newblock   {\it J. Stat. Phys.}  36, 219, 1984. 

 \bibitem {Omn8894} R. Omn\`es, 
\newblock  "Logical Reformulation of Quantum Mechanics: 
I Foundations",
\newblock  {\it J. Stat. Phys.}   53, 893, 1988; 
\newblock  {\it The Interpretation of Quantum Mechanics},
\newblock  (Princeton University Press, Princeton, 1994);
\newblock  "Consistent Interpretations of Quantum Mechanics",
\newblock {\it Rev. Mod. Phys.} 64, 339, 1992. 

  \bibitem {GeHa9093} M. Gell-Mann and J. B. Hartle, 
 \newblock  "Quantum mechanics in the Light of Quantum Cosmology", 
\newblock  in {\it Complexity, Entropy and the Physics of Information}, 
edited by W.\ Zurek,  
\newblock    (Addison Wesley, Reading, 1990);  
\newblock  "Classical Equations for Quantum Systems", 
\newblock {\it Phys. Rev. D}   47, 3345, 1993.  

\bibitem{Har93a} J. B. Hartle,
\newblock "Spacetime Quantum Mechanics and the Quantum Mechanics of Spacetime", 
\newblock in  {\it Proceedings on the 1992 Les Houches School,  	
Gravitation and Quantisation, 1993}.  

\bibitem{Bell64} J. S. Bell, 
\newblock "On the Einstein-Podolsky-Rosen Paradox",
\newblock {\it Physics} 1, 195, 1964.  

 \bibitem{KoSp67} S. Kochen and R. P. Specker, 
\newblock "The Problem of Hidden Variables in Quantum Mechanics",
\newblock {\it J. Math. Mech.} 17, 59, 1967.  

\bibitem{Asp82} A. Aspect, J. Dalibard and  G. Roger,
 \newblock "Experimental Realization of Einstein-Podolsky-Rosen-Bohm Gedankenexperiment: 
a N's Inequalities",
 \newblock {\it Phys. Rev. Lett.} 49, 91, 1982.   

 \bibitem{Bohm52} D. Bohm, 
\newblock "A Suggested Interpretation  of the Quantum Theory in Terms of Hidden variables",
\newblock {\it Phys. Rev.} 85, 166, 1952. 

  \bibitem{BoHi} D. Bohm  and B. J. Hiley, 
\newblock {\it The Undivided Universe},
 \newblock (Routledge, London, 1993). 

 \bibitem{BoAh59} Y. Aharonov and D. Bohm, 
\newblock  "Significance of Electromagnetic Potentials in the Quantum Theory",
\newblock {\it Phys. Rev.} 115, 485, 1959.   

\bibitem{AnSa00} C. Anastopoulos and K. Savvidou, 
 \newblock  "Quantum Mechanical Histories and the Berry Phase",
\newblock quant-ph/0007093. 

\bibitem{Sav99a} K. Savvidou,
\newblock  "The Action Operator in Continuous Time Histories",
\newblock   {\it J. Math. Phys.}  40, 5657, 1999.   

\bibitem{Jau} J. J. Jauch,
\newblock {\it Foundations of Quantum Mechanics},
\newblock (Addison-Wesley, Reading, 1968).

\bibitem{Dav} E. B. Davies,
\newblock  {\it Quantum Theory of Open Systems},
\newblock (Academic Press, London, 1976).


\bibitem{BGL} P. Busch, M. Grabowski and P. J. Lahti,
\newblock {\it Operational Quantum Physics},
\newblock (Springer Verlag, Berlin, 1995). 


\bibitem{Cha60} R. G. Chambers, 
\newblock 
\newblock {\it Phys. Rev. Lett.} 5, 3, 1960. 

\bibitem{SaBh88} J. Samuel and R. Bhandari,
\newblock "General Setting for Berry Phase",
\newblock {\it Phys. Rev. Lett.} 60, 2339, 1988.



 \bibitem{SMP87} D. Suter, K. T. Mueller and A. Pines,
\newblock "Study of the Aharonov-Anandan Quantum Phase by NMR Interferometry",
\newblock {\it Phys. Rev. Lett.} 60, 1218, 1988. 

  \bibitem{DoKe96} F. Dowker and A. Kent, 
\newblock "On the Consistent Histories Approach to Quantum Mechanics",
\newblock {\it J. Stat. Phys.} 82, 1575, 1996.  

\bibitem{Kent97} A. Kent,
 \newblock "Consistent Sets Yield Contradictory Inferences in
 Quantum Theory",
\newblock  {\it Phys. Rev. Lett.} 78,  2874, 1997.  

\bibitem{GriHa98} R. Griffiths and J. B. Hartle,
 "Comment on Consistent Sets Yield Contrary Inferences in Quantum Theory",
\newblock {\it Phys. Rev. Lett.} 81, 1981, 1998.

\bibitem{Gri99}R. Griffiths,
\newblock "Consistent Quantum Counterfactuals",
\newblock  {\it Phys. Rev. A} 60, 5, 1999.

\bibitem{Ish97} C. J. Isham, 
\newblock "Topos Theory and Consistent Histories: the Internal Logic of the Set of All 
Consistent Sets",
 \newblock {\it Int. J. Theor. Phys.} 36, 785, 1997. 

    \bibitem{Sor94} R. D. Sorkin, 
\newblock  "Quantum Mechanics as  Quantum Measure Theory",
\newblock  {\it Mod. Phys. Lett.  A} 9, 3119, 1994.  

\bibitem{Sor97}  R. D. Sorkin, 
\newblock  "Quantum Measure Theory and its Interpretation", 
 in {\it  Quantum Classical Correspondence}, edited by D.H. Feng and B. L. Hu. 
\newblock (International Press, Cambridge MA, 1997).   

\bibitem{An97} C. Anastopoulos,
\newblock "Selection of Preferred Consistent Sets",
\newblock {\it Int. J. Theor. Phys.} 37, 2261, 1998.

\bibitem{I94} C.J. Isham,
\newblock  "Quantum  Logic and the Histories Approach to Quantum Theory",
 \newblock  {\it J. Math. Phys.}  35, 2157, 1994.   

 \bibitem{IL95} C.J.Isham and N. Linden, 
\newblock "Continuous Histories and the History Group in Generalised Quantum Theory",
\newblock  {\it J. Math. Phys.}  36, 5392, 1995.  

 \bibitem{IL94} C.J. Isham and N. Linden,
\newblock "Quantum temporal logic and decoherence functionals in the histories 
approach to generalised quantum theory",
\newblock  {\it J. Math. Phys.}  35, 5452, 1994.   

\bibitem{Nel85} E. Nelson,
\newblock {\it Quantum Fluctuations},
 \newblock (Princeton University Press, Princeton, 1985).  

 \bibitem{Schw61} J. S. Schwinger, 
\newblock  "Brownian Motion of a Quantum Oscillator",
\newblock {\it J. Math. Phys.}  2, 407, 1961.  


\bibitem{Kel64} L. V. Keldysh,
\newblock   "Diagram Technique for Nonequilibrium Processes",
\newblock  {\it Zh. Eksp. Teor. Fiz.}  47, 1515, 1964.



 \bibitem{An00b} C. Anastopoulos, 
\newblock:  "Continuous-time histories: Observables, Probabilities, Phase Space Structure
 and the Classical Limit",
\newblock quant-ph/0008052.


\bibitem{Kla59} J. Klauder. 
\newblock "The Action Option and a Feynman Quantization of Spinor Fields in terms of Ordinary C-
Numbers",
\newblock {\it Ann. Phys} 11, 123, 1959.

 \bibitem{Khre00} A. Khrennikov, 
\newblock  "Einstein and Bell, von Mises and Kolmogorov: Reality and Locality,
 Frequency and Probability",
 \newblock quant-ph/0006016. 

 \bibitem{Stuc} E.C.G. Stueckelberg, 
\newblock  "Quantum Theory in Real Hilbert Space",
\newblock  {\it Helv. Phys. Acta} 33, 727, 1960.   

\bibitem{Kent96} A. Kent, 
\newblock "Quasiclassical Dynamics in a Closed Quantum System",
\newblock {\it Phys.Rev. A} 54,  4670, 1996.  

 \bibitem{GHZ} D. M. Greenberger, M. A. Horne and A. Zeilinger,
 \newblock "Going Beyond Bell's Theorem"   
 \newblock in {\it  Bell's Theorem, Quantum Theory and Conceptions of the  Universe}, 
edited by M . Kafatos.  
\newblock (Kluwer Academica, Dordrecht, 1989). 

 \bibitem{Wood} N. J. Woodhouse, 
\newblock {\it Geometric Quantization.} 
\newblock (Oxford University Press, Oxford, 1992).  

 \bibitem{I83} C. J. Isham,
  \newblock  "Topological and Global Aspects of Quantum Theory" 
\newblock  In {\it Proceedings of the 1983 Les Houches School, Relativity,
 Groups and Topology II}. 

 \bibitem{Aga93} G.S. Agarwal, 
\newblock "Perspective of Einstein-Podolsky-Rosen Spin 
Correlations in the Phase Space Formulation for Arbitrary Values of the Spin",
\newblock {\it Phys. Rev. A} 47, 4608, 1993. 

   \bibitem{ILSS98} C. Isham, N. Linden, K. Savvidou and S. Schreckenberg,
\newblock  "Continuous Time and Consistent Histories",
\newblock   {\it J. Math. Phys.} 37, 2261, 1998.   

 \bibitem{Sav99b} K. Savvidou,
\newblock  "Continuous Time in Consistent Histories",
\newblock gr-qc/9912076.  

  \end{thebibliography}
  \end{document}